\documentclass[preprint,amsmath,amssymb,showpacs, aps,longbibliography]{revtex4-1}

\usepackage[dvips]{graphicx}

\begin{document}

\title{Density of states for systems with multiple order parameters:
a constrained Wang-Landau method}




\author{C.~H. Chan}
\email{seahoi2001@gmail.com }

\author{G. Brown}
\email{gbrown@fsu.edu}

\author{P.~A. Rikvold}
\email{prikvold@fsu.edu}

\affiliation{Department of Physics, Florida State University, Tallahassee, Florida 32306-4350, USA}

\begin{abstract}
A macroscopically constrained Wang-Landau Monte Carlo method was recently proposed
to calculate the joint density of states (DOS) for systems with multiple order parameters.
Here we demonstrate results for a nearest-neighbor Ising  antiferromagnet with
ferromagnetic long-range interactions (a model spin-crossover material).
Its two relevant order parameters are the magnetization $M$ and the staggered
magnetization $M_{\rm s}$.
The joint DOS, $g(E,M,M_{\rm s})$ where $E$ is the total system energy, is
calculated for zero external field and long-range interaction strength, and then
obtained for arbitrary values of these two field-like model parameters by a simple
transformation of $E$. Illustrations are shown for several parameter sets.
\end{abstract}

\maketitle

\section{Motivation, method and model}

To obtain phase diagrams for many-particle systems with multiple order parameters and field-like
system parameters by standard importance sampling Monte Carlo (MC) simulations, one
typically needs to perform simulations for a range of temperatures and field-like parameters.
This is a quite tedious and computationally intensive task. Much effort can
often be saved by calculating
the density of states (DOS) in energy space, $g(E)$, using the Wang Landau (WL)
method \cite{WAN01a,WAN01b} or one of its many later refinements. However, if a {\it joint\/}
DOS involving the energy and several order parameters, $g(E,O_1,O_2,...)$ is needed,
the random walks that lie at the heart of the WL method become multi-dimensional and
consequently quite slow \cite{CHAN17a}.

Motivated by a computational problem of this nature (see specifics below), we recently
developed an algorithm based on the WL method, in which the multidimensional random
walk in energy and order parameters is replaced
by many independent, one-dimensional walks in energy
space, each constrained to fixed values of the order parameters \cite{CHAN17a}.
For this reason, we call it the macroscopically constrained Wang-Landau (MCWL) method.
(A similar, but less general, method was recently proposed by Ren \textit{et al.} \cite{REN16}.)
Our implementation of the algorithm is designed to execute in trivially parallel fashion
on a large set of independent processors.
Through further, symmetry based simplifications \cite{CHAN17a},
the method provides an accurate estimate of a joint DOS with three variables
in a relatively short time.
All the simulations necessary to obtain the main numerical results presented here and in Refs.\
\cite{CHAN17a} and \cite{CHAN17b} were finished within one week, running independently on
about two hundred separate computing cores.
Here, we use the method to obtain the densities of states of a model spin-crossover material
with two macroscopic order parameters,
needed to produce phase diagrams under arbitrary external conditions.
However, it has the potential to be applied to other complicated systems and phase spaces of higher
dimension. Details of the algorithm and its implementation are given in Ref.~\cite{CHAN17a}.

The problem that inspired our development of the MCWL method is a model
spin-crossover material.
These materials are molecular crystals composed of magnetic ions surrounded by
organic ligands. The molecules can be in two different magnetic states,
a highly degenerate, high-spin (HS) and a less degenerate, low-spin (LS) state.
The HS state has higher energy and larger molecular volume than the LS state, and the volume difference leads to elastic distortions of crystals with a mixture of HS and LS molecules.
A simple model of phase transitions in some spin-crossover materials can be constructed as an
antiferromagnetic Ising-like model with elastic distortions \cite{NISH13}.
The elastic interactions can be further
approximated as a ferromagnetic mean-field (equivalent neighbor or Husimi-Temperley)
interaction, leading to the following Hamiltonian on an $L \times L$
square lattice with periodic boundary conditions \cite{CHAN17a,BROWN201420,RIKV16},
\begin{equation}\label{def_Hamiltonian_2D_Ising-ASFL}
\mathcal{H} = J \sum_{\langle i,j \rangle}s_{i}s_{j}-HM -\frac{A}{2L^{2}}M^{2} ~,
\end{equation}
with $J>0$. Each spin, $s_{i}$ is either up ($+1$, HS) or down ($-1$, LS).
$J \sum_{\langle i,j \rangle}s_{i}s_{j}$ represents the short-range interaction between neighboring spins, which favors adjacent spins aligning alternately up and down in an
antiferromagnetic (AFM) configuration.
$M=\sum_{i}s_{i}$ is the magnetization and $H$ is the strength of the
applied field (which is actually an effective field in the spin-crossover material
\cite{RIKV16,WAJN71}).
The term $-HM$ favors all spins
aligned with $H$ in the same direction. $L^{2}$ is the total number of sites, and $A\geq0$ is the long-range interaction strength. Two spins located far apart on the lattice will interact through the term $M^{2}$, such that  $-\frac{A}{2L^{2}}M^{2}$ has its most negative value if all the spins are aligned in the same direction. Therefore, this amounts to a ferromagnetic (FM)
long-range interaction.
We will refer to this model as the two-dimensional Ising Antiferromagnetic Short-range and
Ferromagnetic Long-range (2D Ising-ASFL) model.
Its two field-like system parameters are $H$ and $A$.

Throughout the paper, the total energy ($E$), magnetic field ($H$), and long-range interaction
($A$),  will be expressed in units of $J$.
The lattice is divided into two interpenetrating sublattices,
A and B in a checkerboard fashion, with sublattice magnetizations $M_{\rm A}$ and $M_{\rm B}$,
respectively.
The two order parameters of the model are the magnetization ($M = M_{\rm A} + M_{\rm B}$)
and the staggered magnetization ($M_{\rm s} = M_{\rm A} - M_{\rm B}$).

The conditional DOS, $g(E | M,M_{\rm s})$, is determined through independent,
one-dimensional WL simulations for fixed $M$ and $M_{\rm s}$.
The order parameters are kept constant by performing microscopic spin-exchange
(Kawaski) updates separately on each sublattice. The joint DOS is then obtained
by an analogue of Bayes' theorem as
\begin{equation}\label{def_joint_DOS}
g(E,M,M_{\rm s})=\frac{g(E|M,M_{\rm s})}{\sum_{E}g(E|M,M_{\rm s})}g(M,M_{\rm s}) \;,
\end{equation}
where $g(M,M_{\rm s})$ is determined by exact combinatorial calculation.
It is only necessary to calculate $g(E | M,M_{\rm s})$ for one single set of $H$ and $A$.
Here we use  $H=A=0$, which corresponds to a simple square-lattice Ising
antiferromagnet \cite{LOUR16}.
The joint DOS for any arbitrary value of $(H,A)$ can be obtained by
the simple transformation of $E$,
\begin{equation}
\label{def_shift_E_DOS}
g(E,M,M_{\rm s}) \rightarrow g(E-HM- \frac{AM^{2}}{2L^{2}},M,M_{\rm s}) \;.
\end{equation}
This result is based on the fact that all microstates are equally shifted in energy
when a field-like parameter, such as $H$ or $A$, couples to a function of a global property,
such as $M$ (or $M_{\rm s}$), as shown in  Eq.~(\ref{def_Hamiltonian_2D_Ising-ASFL}).

\section{Results}
\begin{figure}
\begin{center}
\includegraphics[width=37pc]{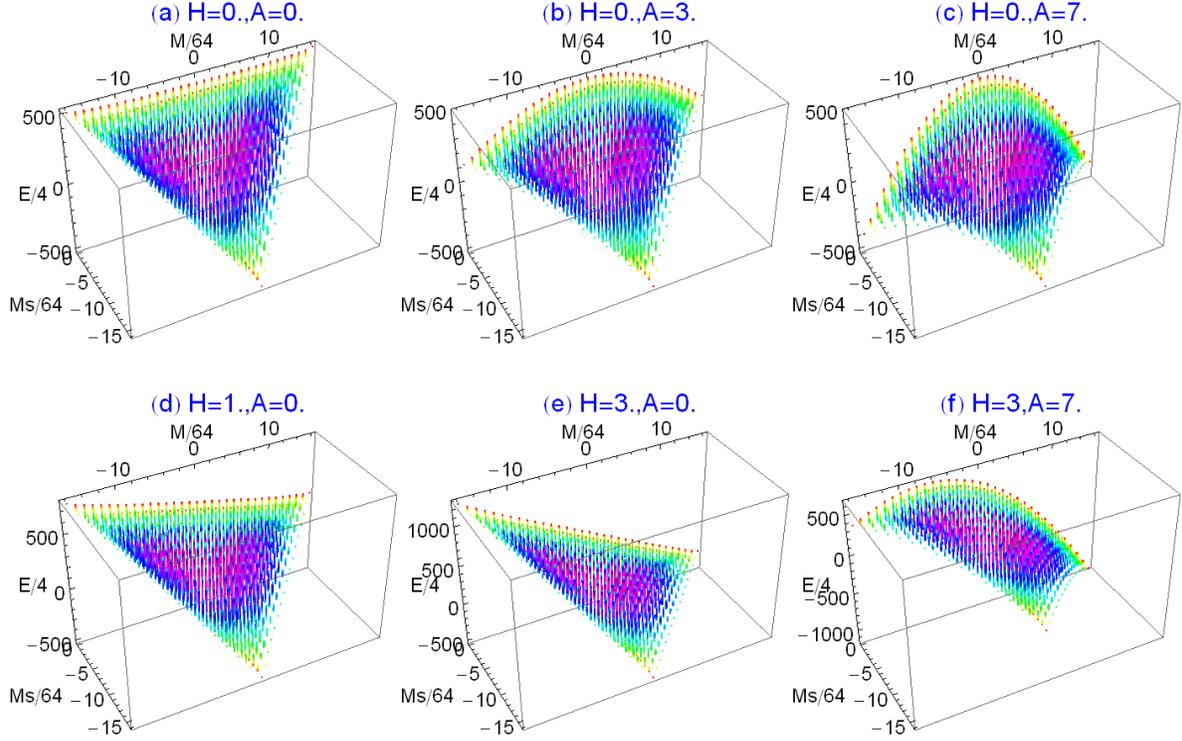}
\end{center}
\caption{Natural logarithm of the joint DOS, $\ln g(E,M,M_{\rm s})$,
of the 2D Ising-ASFL model for system size $L=32$ for different field strengths ($H$) and
long-range interaction strengths ($A$). The three axes are $M/64$,
$M_{\rm s}/64$, and $E/4$.
The colors of the data points show the relative magnitude of the natural logarithm of the DOS,
ranging from red (lowest) to magenta (highest). Only results for $M_{\rm s}\leq 0$
are shown as there is reflection symmetry about the $M_{\rm s} = 0$ plane.
$M$ and $M_{\rm s}$ are both sampled with an increment of 64. All possible energy levels are
included, giving an increment of 4 for $E$ at fixed $M$ and $M_{\rm s}$.
}\label{WLfig_allLnDOS8a_L6}
\end{figure}

\begin{figure}
\begin{center}
\includegraphics[width=1.0\textwidth]{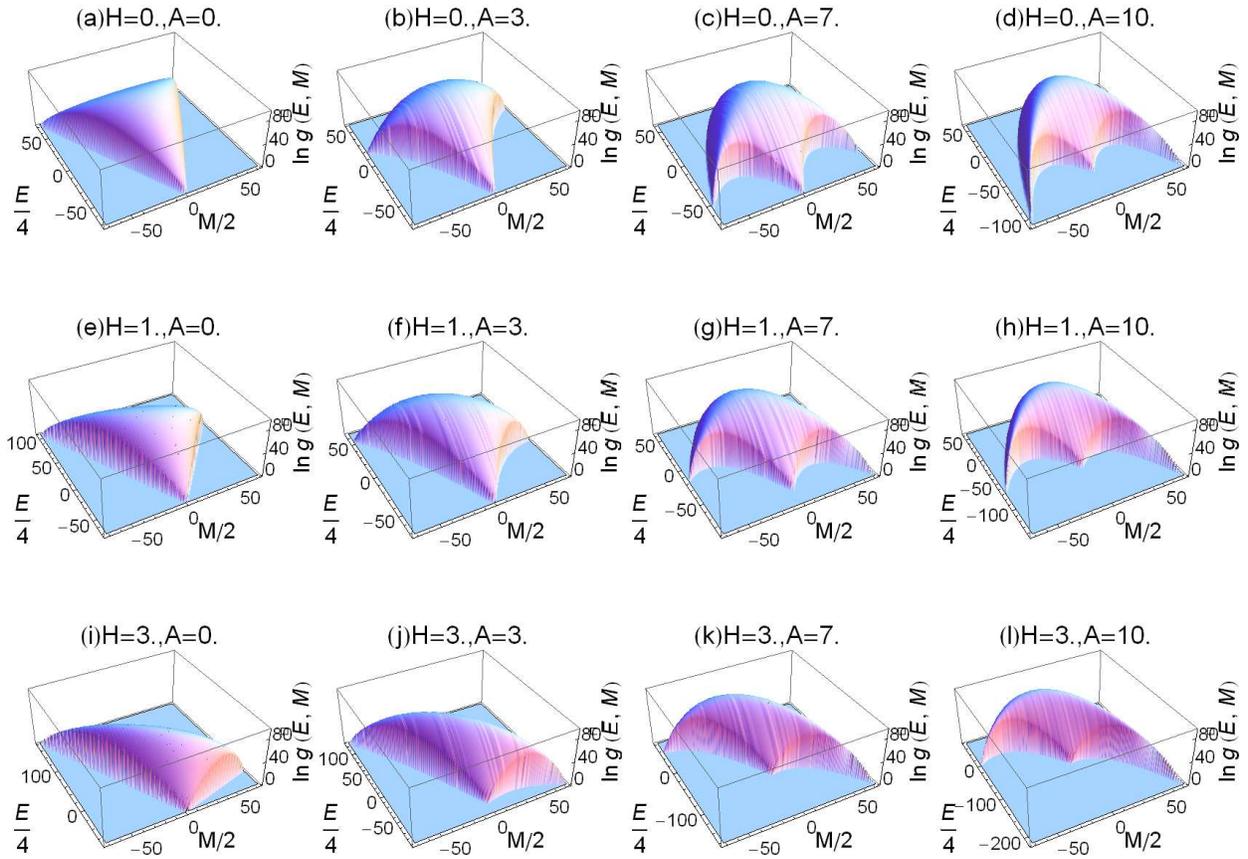}
\end{center}\caption{Plots of the logarithmic DOS vs energy ($E$) and magnetization ($M$),
$\ln g(E,M)$, for different fields $H$ and long-range interactions $A$.
The smaller system size, $L=12$, is used for better graphical resolution.
All possible values of $E$ and $M$ for this system size are considered.
}\label{WL_fig_DOS_vs_ME}
\end{figure}

\begin{figure}
\begin{center}\includegraphics[width=1.0\textwidth]{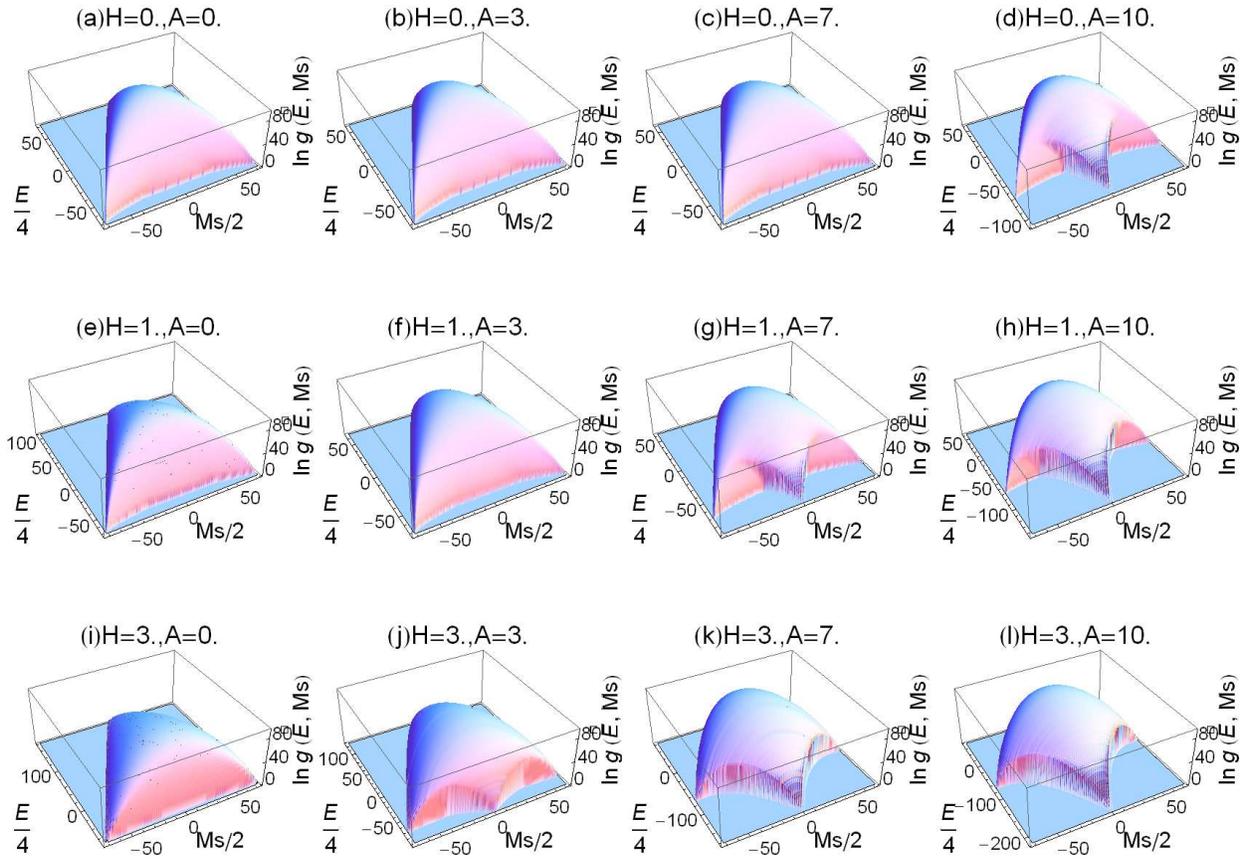}
\end{center}
\caption{Plots of the logarithmic DOS
vs energy ($E$) and staggered magnetization ($M_{\rm s}$), $\ln g(E,M_{\rm s})$,
for different fields $H$ and
long-range interactions $A$, again using $L=12$. All possible values of $E$ and $M_{\rm s}$  for
this system size are considered.
}\label{WL_fig_DOS_vs_MsE}
\end{figure}

Figure \ref{WLfig_allLnDOS8a_L6} shows the joint  DOS,  $g(E,M,M_{s})$, for different
magnetic fields $H$ and long-range interaction strengths $A$.
Simulations were only performed for the case $H=0$ and $A=0$,
shown in (a).
Only $1/8$ of the data points in (a) need to be calculated directly by simulation.
The remaining data points are obtained through symmetry transformations in the
$(M,M_{\rm s})$ plane as described in \cite{CHAN17a}.
Data points in (b) to (f) are all obtained through transformation of the energy axis in (a),
using Eq.~(\ref{def_shift_E_DOS}). When the field $H$ is applied to the system as in (d) and (e), the term $-HM$ in the Hamiltonian pulls down the states with positive magnetization ($M>0$) linearly, while it pulls up the states with negative magnetization ($M<0$). On the other hand, with the long-range interaction ($A$) present, due to the term $-AM^2/2L^2$, states with non-zero magnetization ($M$) are pulled down in a quadratic manner. If both $H$ and $A$ are applied, some states are shifted up, and some are shifted down as shown in (f). Figure
\ref{WLfig_allLnDOS8a_L6} shows results for the largest system size we studied, $L=32$.
While Fig.~\ref{WLfig_allLnDOS8a_L6}  shows the DOS versus three variables ($E,M,M_{\rm s}$),
Figs.~\ref{WL_fig_DOS_vs_ME} and \ref{WL_fig_DOS_vs_MsE} show the DOS
after summing the results along one axis, i.e., when the DOS
is plotted against two variables only, $E$ and $M$, and $E$ and $M_{\rm s}$, respectively.

Figure \ref{WL_fig_DOS_vs_ME} shows $g(E,M)$.
A similar effect as discussed for Fig.~\ref{WLfig_allLnDOS8a_L6} is seen.
I.e., when $H$ is applied, the energies for states with positive $M$
are pulled down, and the energies for states with negative $M$ are pulled up.
When $A$ is applied, the energies for all states with non-zero $M$ are pulled down quadratically.

Figure \ref{WL_fig_DOS_vs_MsE} shows $g(E,M_{\rm s})$.
Parts (a), (e), and (i) show the effect of increasing $H$ with $A=0$.
The energy range of states with $M_{\rm s}$ near zero increases as states with negative $M$
are pulled up in energy, and states with positive $M$ are pulled down. The latter effect also causes
the energy corresponding to the highest DOS to shift to a lower value.
The other graphs in Fig.~\ref{WL_fig_DOS_vs_MsE} show that increasing the long-range
interaction strength $A$ lowers the minimum energy of states with $M_{\rm s}$ near zero, as the
energies of the corresponding
states with large positive and negative $M$ are pulled down quadratically.

\section{Summary}
As an example of the application of the recently proposed macroscopically constrained
Wang-Landau Monte Carlo method for systems with multiple order parameters,
we have presented the joint DOS of a square-lattice Ising model with antiferromagnetic short-range
interactions and ferromagnetic long-range interactions
(a model spin-crossover material \cite{RIKV16}).
Based on data from simulation for one single parameter set,
results are shown for various values of the model's field-like parameters, applied field $H$ and
long-range interaction strength, $A$.
Further results on the phase diagrams of this
model spin-crossover material are given in Ref.\ \cite{CHAN17b}.


\section*{ACKNOWLEDGMENTS}

Chan thanks Alexandra Valentim and Ying Wai Li for helpful discussions of the
WL method. The Ising-ASFL model was first proposed by Seiji Miyashita,
and we thank him for useful discussions.
The simulations were performed at the Florida State University High Performance Computing
Center.
This work was supported in part by NSF grant No. DMR-1104829.



\end{document}